\documentclass[journal,12pt,onecolumn,draftclsnofoot,]{IEEEtran}
\usepackage[T1]{fontenc}
\usepackage{amsmath}
\usepackage{cite}
\usepackage{setspace}
\usepackage{verbatim}
\usepackage{url}
\usepackage{graphicx}
\usepackage{color}
\usepackage[justification=centering]{caption}
\usepackage{subcaption}
\usepackage{float} 
\usepackage{hyperref}
\usepackage{silence}
\usepackage{etoolbox} 
\newbool{DEBUG}
\ifbool{DEBUG}{\linespread{0.9}}

\begin{document}
\title{Control and Management of Multiple RATs in Wireless Networks: An SDN Approach}
\author{Akshatha Nayak M., Arghyadip Roy, Pranav Jha, and Abhay Karandikar

\thanks{\IEEEcompsocthanksitem \textit{The authors are with the Department
of Electrical Engineering, Indian Institute of Technology Bombay,
Mumbai, 400076, India. e-mail: {$\lbrace$akshatha, arghyadip, pranavjha, karandi$\rbrace$}@ee.iitb.ac.in.
Abhay Karandikar is currently the Director, Indian Institute of Technology Kanpur (on leave from IIT Bombay), Kanpur, 208016, India. e-mail:karandi@iitk.ac.in.}}}

\maketitle

\begin{abstract}
Telecom operators are using a variety of Radio Access Technologies (RATs) for providing services to mobile subscribers. This development has emphasized the requirement for unified control and management of diverse RATs. Although multiple RATs co-exist within today's cellular networks, each RAT is controlled by a set of different entities. This may lead to suboptimal utilization of the overall network resources. In this article, we review various architectures for multi-RAT control proposed by both industry and academia. We also propose a novel SDN based network architecture for end-to-end control and management of diverse RATs. The architecture is scalable and provides a framework for improved network performance over the present day architecture and proposals in existing literature. Our architecture also provides a framework for deployment of applications in a RAT agnostic fashion. It facilitates network slicing and enables the provision of Quality of Service (QoS) guarantees to the end user. We have also developed an evaluation platform based on ns-3 to evaluate the performance offered by the architecture. Experimental results obtained using the platform demonstrate the benefits provided by our architecture.
\end{abstract}

\section*{Introduction}
The widespread deployment of Fourth Generation (4G) cellular networks has brought about a radical change in the nature of mobile data consumption. Data-intensive mobile applications such as social networking, video streaming, online-gaming etc., are becoming increasingly popular. This phenomenon, along with a growing subscriber base has created the need for higher capacity networks. To cater to increasing data traffic demands, mobile network operators are supplementing cellular network deployments with Wireless Local Area Networks (WLANs). Plans for deployment of the next generation cellular networks known as the Fifth Generation (5G) networks as early as 2020, are also underway. As a result, the complexity of managing the mobile network is continually increasing with newer Radio Access Technologies (RATs) being added over time. 

\par A multitude of RATs exist in today's wireless networks and each of the these RATs is controlled by one or more RAT specific entities e.g., the Mobility Management Entity (MME), Policy and Charging Rules Function (PCRF) etc., in the Long Term Evolution (LTE) network and WLAN controller in the WLAN, resulting in a fragmented control of multi-RAT wireless networks. The fragmentation of the control plane prohibits a global view of the network resources. Although solutions for unified control of multiple RATs have been proposed by academia, they have not yet been deployed. Even in the upcoming Third Generation Partnership Project (3GPP) 5G network which supports multiple RATs, radio access related decisions are taken separately within an access network for various RATs~\cite{5gspec}. Also, the core network and the Radio Access Network (RAN) are managed separately. By devising RAT agnostic control for applications that are common to all wireless networks e.g., User Equipment (UE) authentication, mobility management and flow control, we can control and manage diverse RATs in a unified manner. One of the possible ways of achieving this objective is with the aid of the Software Defined Networking (SDN) paradigm.

\par SDN~\cite{sdnrfc} is a networking principle that decouples the control and data planes. The control plane of a network comprises of control and management elements and protocols, whereas the data plane comprises of elements/functions that forward data. The two planes are separated by a standardized interface. This interface facilitates the configuration of data plane elements using policy-based rules and eliminates the need for vendor-specific configurations. This interface can also expose the capabilities of network elements which could be used by third-party vendors for developing applications. 

\par In this article, we present an SDN based wireless network architecture with well defined separation of control and data planes, which unifies the control and management of diverse RATs. The architecture consists of a slice manager which splits the end-to-end physical network into multiple logical networks or network slices. Each slice comprises of data plane nodes and a control plane entity known as the multi-RAT controller to manage the nodes in a unified manner. Our architecture provides a framework for the deployment of RAT agnostic control applications. Additionally, it provides the flexibility to support other future RATs in the integrated framework. Usage of network slicing with a controller for each network slice also brings scalability to the architecture. 

\par The contributions offered by the proposed architecture are as follows:
\begin{itemize}
\item The architecture provides end-to-end control of the multi-RAT network while ensuring scalability. It ensures a global view of the network at the controller which allows for improved decision making for applications.
\item As our solution uses a single controller for managing the core and the RAN, the control signaling between the core network controller and RAN controllers present in existing multi-tiered approaches is no longer required. 
\item It provides a RAT agnostic interface for control applications making application development simpler.
\item The signaling towards the UE remains broadly unchanged, making this architecture ideal for practical deployments contrary to existing approaches.
\item The architecture also supports network slicing, which not only ensures Quality of Service (QoS) but also provides scalability to the network.
\end{itemize} 

\par The rest of the article is organized as follows. The next section describes the existing work on control and management of multi-RAT networks. The details of the proposed SDN based architecture to control and manage multi-RAT networks are provided in the succeeding section. Advantages of this architecture are described in the subsequent section. A few experimental results are then provided, followed by conclusion.

\section*{Architectures for Multi-RAT Network Control}
The 5G network proposed by 3GPP consists of multiple RATs. The 3GPP 5G network not only supports 3GPP LTE and New Radio (NR) technology but also non-3GPP RATs such as WLAN. The 3GPP 5G network introduces the usage of SDN for control and management but both the RAN and the core network have separate control architectures. Existing works such as~\cite{akyildiz2015softair, zhang2017network,ali2013crowd,rahman2015hetnet} present two tiered cloud architectures for control of multi-RAT networks. Control and management tasks related to mobility, resource allocation and interference are handled by the core cloud, whereas the edge cloud (which is closer to the UEs takes care of the RAN functions. The lower-tier controller carries out tasks that occur more frequently whereas the controller on the upper-tier works on less frequent activities. Authors in~\cite{rahman2015hetnet} utilize the spare bits of the OpenFlow packet model to implement virtual networks and enable multi-RAT control. This architecture makes  use of a higher level network controller for provisioning network nodes and local controllers at the remote radio heads. Works such as \cite {munasinghe2017traffic,mi2014no,fengyi2015flexible} have proposed three tiered architectures for multi-RAT control. A three-tiered architecture consisting of the physical, control and management layers for dense multi-RAT networks has been proposed in~\cite{munasinghe2017traffic}. In~\cite{mi2014no}, the authors present a three tiered architecture with a flat user plane. This is achieved by encapsulating the protocol layers of the controlled RATs as a module. An increase in the granularity of modules brings about an improvement in flexibility but results in an additional management cost. In~\cite{fengyi2015flexible}, an architecture comprised of three clouds, based on the functionality of the network elements viz., control, access and forwarding clouds is described.This architecture is realized using Network Function Virtualization (NFV) and service improvement is achieved by placing the user plane functions e.g., Gateways (GWs) closer to the network edge. An approach for unified control and management of multi-RAT networks has been described in~\cite{kitindi2017wireless} which proposes a clean slate architecture known as Cloud-RAN (C-RAN). Within C-RAN, most of the network processing including the baseband function processing, is carried out in the cloud. 

\par Existing works on slicing in the multi-RAT scenario include~\cite{rost2017network} and \cite{ferrus20185g}. In~\cite{rost2017network}, slicing is achieved by aggregating the network entities that are shared by different services into common sub-slices which are controlled by a co-ordinator. This entity ensures individual service QoS by co-ordinating and prioritizing across functions.This solution also consists of dedicated network entities for other services and is administered by another controller entity. Authors of~\cite{ferrus20185g} propose a framework to specify and support creation of RAN slices using configuration descriptors at every layer of the radio protocol stack, i.e., L1, L2, L3 layers of the 5G stack. These descriptors are used to characterize the policies, features and resources within the protocol layers. 
\par Although a few works in the existing literature have focused on solutions for multi-RAT control, to the best of our knowledge, no other work presents a unified SDN based framework for end-to-end network control while ensuring scalability through the creation of multiple logical networks over a single physical network. Our work also abstracts out the RAT specific details from the application to enable a uniform method for control and management of multi-RAT wireless networks.

\section*{SDN based Multi-RAT Network Architecture} 
In this section, we present an architecture that provides a unified framework to support multiple RATs using the principles of SDN. The architecture comprises of two types of control entities viz. multi-RAT controllers and the slice manager.
The slice manager is an entity that creates logical resource units by grouping some of the physical resources within the network based on service (or load) requirements and isolates them from one another. The set of these logical resource blocks taken together end-to-end are known as slices. The slice size can be increased or shrunk by re-grouping the physical resources. Each slice consists of data plane functions such as data plane Base Stations (dBSs), GWs, cache server etc., and control plane entity viz. the multi-RAT SDN controller. 
\par
\begin{figure}
	\centering
		\includegraphics[width=\linewidth]{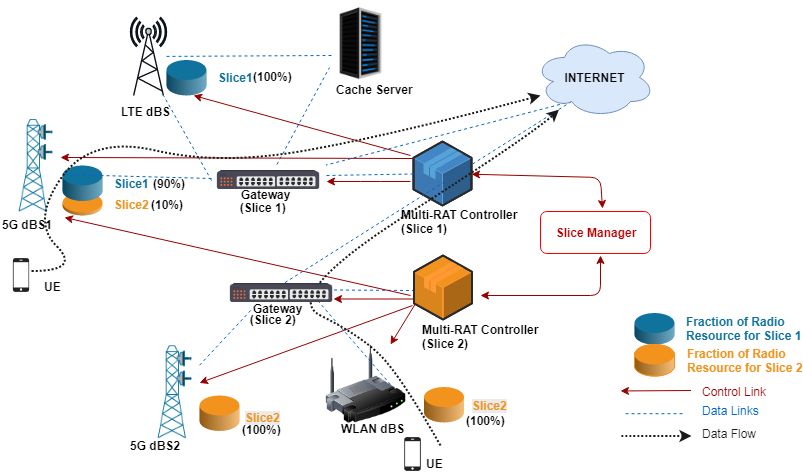}
		\caption{SDN based multi-RAT network architecture.}
        \label{fig:deployment}
 \end{figure}

 The controller controls and manages the data plane entities within the slices and provides data flow configurations to them. It is also responsible for exchanging control plane messages with the UEs. It may also exchange control plane messages with controllers which are a part of other slices. In an example scenario, depicted in Figure~\ref{fig:deployment}, each network slice has a controller which controls the dBSs and GWs belonging to the slice. Controllers may also be shared across slices. Each slice may be governed by specific policies for resource management. Slice-specific policies enable provisioning of services with specific QoS requirements. The proposed architecture provides scalability due to the presence of multiple slices and controllers.
\par The dBSs are RAT specific data plane entities which are created by eliminating the control functionality such as radio resource management, mobility management etc., from the respective RAT specific base stations. For example, as illustrated in Figure~\ref{fig:dbsprotocolstack}, an LTE dBS and a 5G dBS consists of only the lower layers of the stack viz., Packet Data Convergence Protocol (PDCP), Radio Link Control (RLC), Medium Access Control (MAC) and Physical Layer (PHY). The Radio Resource Control (RRC) layer which consists of control functionality is eliminated from the dBS. Its functionality is incorporated into the controller. Similarly WLAN dBSs consists of only PHY and lower MAC layers. dBSs are also responsible for forwarding signaling/control plane messages that are exchanged between the UEs and the controller. They are also responsible for forwarding user plane data exchanged between UEs and external data networks either directly or via the GWs. GWs are generic data plane nodes, which are responsible for forwarding user plane data towards other GWs or external data networks. A GW supports data forwarding for all types of UEs and all types of RATs. 

\subsection*{Multi-RAT Controller Architecture}

\begin{figure}
\caption{Data and control plane nodes in the SDN based multi-RAT architecture with (a) data plane node and (b) SDN based multi-RAT controller.}
   \begin{subfigure}{\textwidth}
		\includegraphics[width=\linewidth]{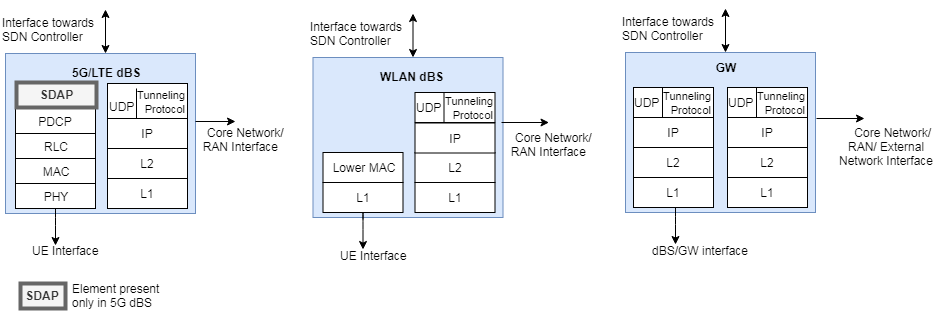}
		\caption{}
       \label{fig:dbsprotocolstack}
	\end{subfigure}
	\begin{subfigure}{\textwidth}
		\includegraphics[width=\linewidth]{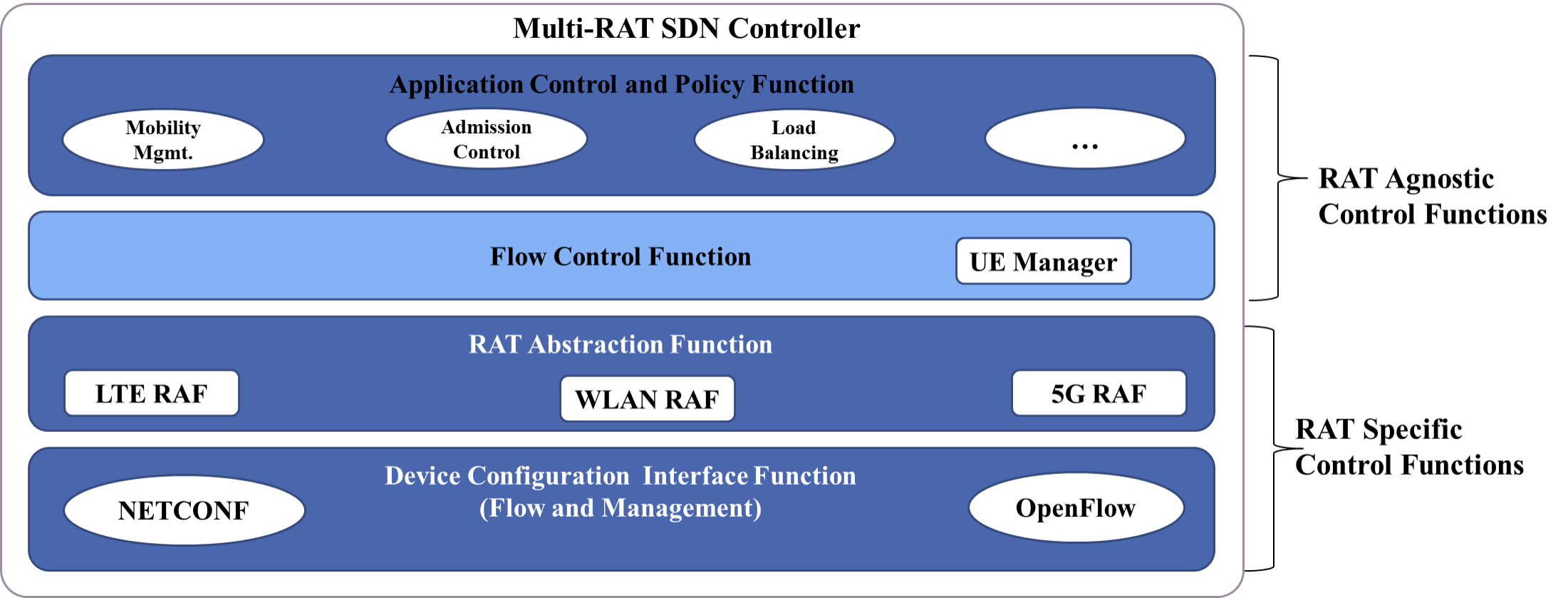}
		\caption{}
       \label{fig:proposedmultiratdep}
	\end{subfigure}
\end{figure}

\par Figure~\ref{fig:proposedmultiratdep} illustrates the architecture of the multi-RAT SDN controller.  In order to control multiple RATs in a unified manner, functions such as UE authentication, UE mobility  management and flow control could be handled in a RAT agnostic manner. As a result, the controller comprises of functionality providing RAT agnostic control and RAT-specific control. The controller comprises of various functions, viz., : 
\begin{enumerate}
\item Device Configuration Interface Function (DCIF): DCIF is the lowest layer of the controller. It interfaces with the data plane functions, i.e., dBSs and GWs through management and control protocols, e.g., NETCONF and OpenFlow~\cite{nunes2014survey}. DCIF is utilized by the controller to configure the data plane functions.
\item RAT Abstraction Function (RAF): This function is responsible for handling the RAT specific functionality within the network. There may exist a separate RAF for every supported RAT. It also manages RAT specific control plane communication with UEs. The function possesses both management and control functionality and is used to translate generic configuration provided by higher layer functions into RAT specific configuration to be supplied to a dBS via the DCIF. For example, the 3GPP LTE RAF translates generic flow configuration parameters provided by the layer above into radio bearer parameters to be supplied to an LTE dBS. It also manages RAT specific control plane communication with UEs. The DCIF and the RAF operate on data that is RAT specific. The rest of the modules are RAT agnostic.
\item Flow Control Function (FCF): FCF deals with an abstract view of the underlying network. The function is responsible for setting up flows on dBSs and GWs with the desired QoS requirements. It also provides a RAT-independent interface to the layer above which may contain RAT agnostic control algorithms. FCF maintains a unified list of abstract attributes for each connected UE and its associated data flows.  It also contains a UE Manager module to store information related to the UE context. 
\item Application Control and Policy Function (ACPF): ACPF comprises of slice-specific control and policy applications. Operators can introduce new applications/policies into a specific slice without affecting other network slices. A RAT independent interface between ACPF and the FCF enables third-party vendors to implement new algorithms without the necessity of understanding the underlying network complexity. 
\end{enumerate}

\subsection*{Control and Management Procedures in the SDN based Multi-RAT Network}
This section describes some of control and management procedures in our network architecture. LTE and WLAN have been used as reference RATs to describe different procedures related to the architecture. These procedures can also be extended to incorporate other RATs, such as the 5G NR RAT.

\begin{itemize}
\item \textbf{UE Association:} Figure~\ref{fig:genericassociation} illustrates UE association call flow in LTE RAT in the architecture. In this network, control messages such as RRC Connection Request are forwarded to the controller for processing. These message are encapsulated within OpenFlow messages. Within the controller, the RAF decodes this message and sends an Admission Request (a RAT independent  message) to the FCF. This message is then forwarded to the admission control application. RAF responds with an RRC Connection Setup message to the UE based on the response from the application and creates a signaling radio bearer between the LTE dBS and the UE. On receipt of this message, the UE sends an Attach Request to the controller utilizing the newly created signaling radio bearer. The RAF receives this message and initiates the authentication/identity procedures. Following this, RAF initiates the creation of a default data bearer between the UE and the dBS and also sends the Attach Accept message to the UE. UE may then initiate data transfer over the default bearer. In the absence of a matching rule at the dBS for handling the received data packets (flow) over the newly created bearer, the data packets are forwarded by the dBS to the controller. The controller may analyze the packet and a dedicated bearer may also be setup from the dBS to the GW to complete the data path through the wireless network. 

\begin{figure}
\caption{Call flows in the SDN based multi-RAT architecture with (a) UE association call flow in LTE and (b) UE handover call flow from WLAN to LTE.}
	\centering
	\begin{subfigure}{\textwidth}
		\includegraphics[width=0.8\linewidth]{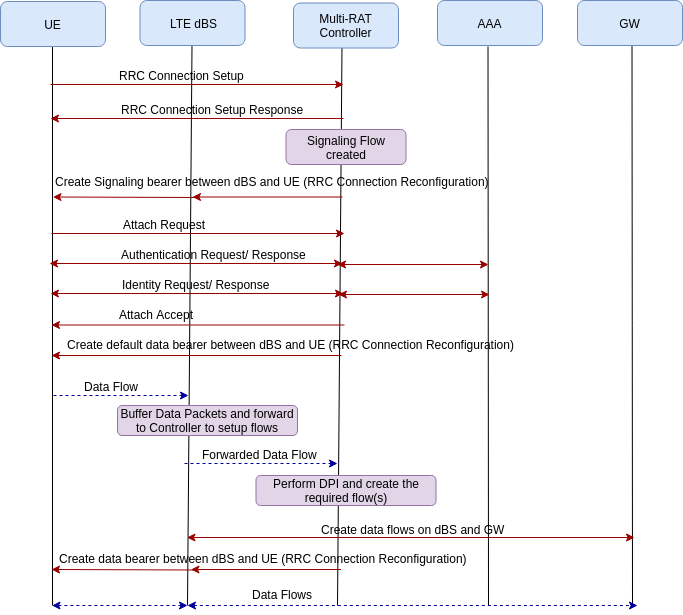}
		\caption{}
        \label{fig:genericassociation}
	\end{subfigure}

	\begin{subfigure}{\textwidth}
		\includegraphics[width=0.8\linewidth]{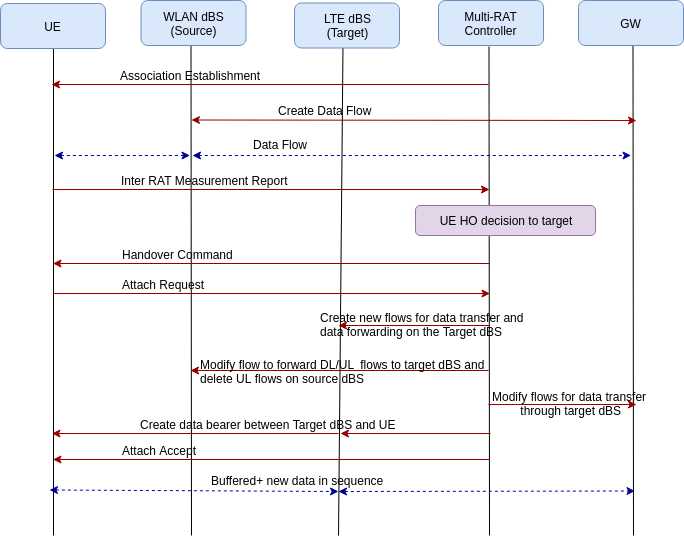}
		\caption{}
        \label{fig:generichandover}
	\end{subfigure}

\end{figure}

\item{\textbf{Mobility Management:}} User mobility is managed in a unified manner in this architecture. The decision to perform handover for a UE is taken by the mobility management function of the ACPF within the controller. We illustrate the unified mobility management process with the help of an inter-RAT mobility call flow (WLAN to LTE) in Figure~\ref{fig:generichandover}. The protocol stack processing for messages is similar to the previous example. The measurement reports from the UE are forwarded to the controller to assist in the handover decision. After the handover, UE is associated with an LTE dBS. Since the UE context is maintained at the controller, re-authentication may not be required. Also, the decision making at multiple individual nodes, such as the source and target dBSs, as done in the existing wireless networks, is no longer needed.
\end{itemize}

\section*{Advantages of the SDN based Multi-RAT Network Architecture}
The proposed network architecture offers multiple advantages in comparison to present day network architecture and existing architectures in literature as described below. 
\begin{itemize}
\item{\textbf{Unified Authentication and Security:}} The authentication and security procedures are handled by the controller. Authentication, which is carried out in a unified manner, prevents the need for authenticating the UE every time it connects to a different RAT. This also enables seamless handovers.  
\item \textbf{Simplified Signaling Procedures:} The signaling procedures are simplified due to unified control. Messages with a request-response format, which are required in existing wireless networks, are reduced due to a unified framework for decision making. Also additional signaling messages which are present within multi-tier networks for maintaining state consistency are no longer needed.
\item{\textbf{Support for multi-connectivity:}} The architecture enables UEs to be attached to multiple RATs at the same time. Due to the presence of a logically unified multi-RAT controller, the control plane interaction between different network entities becomes much simpler compared to the dual connectivity mechanisms in the existing network.
 \item \textbf{Increased flexibility:} The architecture provides flexibility within the network in multiple ways. The radio coverage of the network can be increased by introducing new nodes or configuring existing nodes (which have multiple radio interfaces) as relays within the network. It also enables provisioning of newer services by creating specific slices and can be adapted to incorporate future RATs. As the data path is configured by the controller, service function chaining can be performed in a dynamic manner. The architecture allows for a flexible distribution of protocol layers across data plane entities and the configuration shown in Figure~\ref{fig:dbsprotocolstack} is just one of the many possible configurations. 
\end{itemize}
\par The network architecture also provides a global view of the network to the controller. This allows for the implementation of better control and management algorithms. Some of the areas where improved results may be obtained are:
\begin{itemize}
\item{\textit{Energy efficiency and power control:}} Unlike in present day multi-RAT networks, the SDN controller can regulate power levels for the entire system, thus reducing the overall interference in the RAN. This unified interference management may result in better system throughput. Some dBSs can even be turned off during periods of low traffic by re-distributing the load to the active base stations for increased energy saving. 
 \item{\textit{Content caching and delivery:}} By inspecting packets at the controller, data request for popular content can be retrieved from locations near the dBSs instead of the external network through the GW. This results in reduced content retrieval time as well as efficient backhaul usage. Additionally, the source dBS may itself act like an anchor point and continue to serve the UE even after its handover to another dBS.
 \item{\textit{Radio Resource Management (RRM):}} Due to an abstract view of the radio resources provided by the RAF to the application, RRM procedures such as load balancing, resource allocation etc., can be performed more efficiently in comparison to the existing networks. 
 \end{itemize}

\section*{Experimental Results}
In this section, we characterize improvements offered by the SDN based multi-RAT architecture in comparison to present day wireless networks. We have developed an ns-3 based evaluation platform in harmonization with the proposed architecture, for measuring the system performance. This platform modifies the functionality of LTE and WLAN base stations in ns-3 and converts them into pure data plane nodes. The platform also comprises of a multi-RAT controller module which aggregates the control functionality of the RATs and can configure the data path according to user defined policy.

\par The architecture provides a global view of the network. This is unlike the existing multi-RAT network architecture which provides network information only within a given RAT. The global view provided by the SDN based multi-RAT architecture enables us to implement algorithms which operate on the global network view. We demonstrate the performance improvements provided by our architecture vis-a-vis traditional network architectures using two scenarios. In the first scenario, we consider a system with an LTE BS and a WLAN Access Point (AP) with overlapping coverage areas. We measure the throughput and latency for best effort data traffic in the above system using traditional network selection schemes. We also evaluate the same metrics within our architecture using a simple heuristic which uses the global network view. The simulation parameters for WLAN and LTE as illustrated in Table~\ref{table:lte}, have been obtained from~\cite{hetnetspec}.

\begin{table}
\caption{Multi-RAT network model for LTE and WLAN.}\label{table:lte}
\centering 
\begin{tabular}{|l||l|}
\hline
\textbf{Parameter} & \textbf{Value} \\ \hline
Data rate for a single LTE user & $5$ Mbps \\ \hline
Mean service time for user & $60$ s \\ \hline
Path loss & $128.1+37.6 \log(R)$,  $R$ in kms \\ \hline
WLAN channel bit rate & $54$ Mbps \\ \hline
Tx power for LTE (d)BS & $46$ dBm \\ \hline
Tx power for AP and WLAN dBS & $23$ dBm \\ \hline
Tx power for  UE & $23$ dBm \\ \hline
\end{tabular}
\end{table}

Due to the absence of a global view of the network, it is difficult to achieve load balancing across WLAN and LTE RATs in present day networks. However, in the SDN based multi-RAT network, the information of load at both WLAN and LTE dBSs are available to the controller, making centralized load balancing possible. We implement the following simple load based heuristic in the controller wherein arriving UEs are associated with the WLAN dBS upto a certain threshold. All other UEs are preferably associated with the LTE dBS until the LTE system capacity is reached.

\begin{figure}
\caption{Simulation results for (a) system throughput v/s $\lambda_d$ for scenario I, (b) data transfer latency v/s $\lambda_d$ for scenario I, (c) data slice throughput v/s $\lambda_d$ for scenario II and (d) video user blocking probability v/s $\lambda_v$ for scenario II.}
\label{fig:simulation_results}
	\centering
	\begin{subfigure}{0.45\textwidth}
		\includegraphics{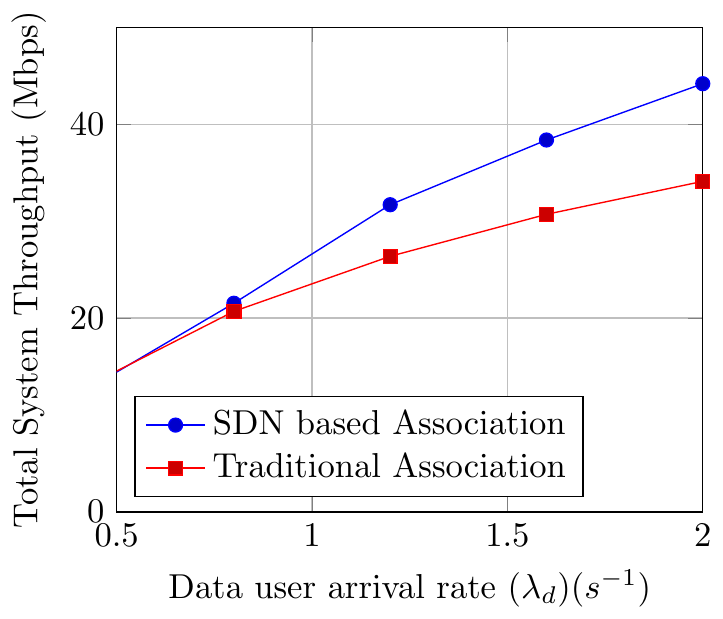}
		\caption{}
        \label{fig:system_throughput}
	\end{subfigure}
	\begin{subfigure}{0.45\textwidth}
		\includegraphics{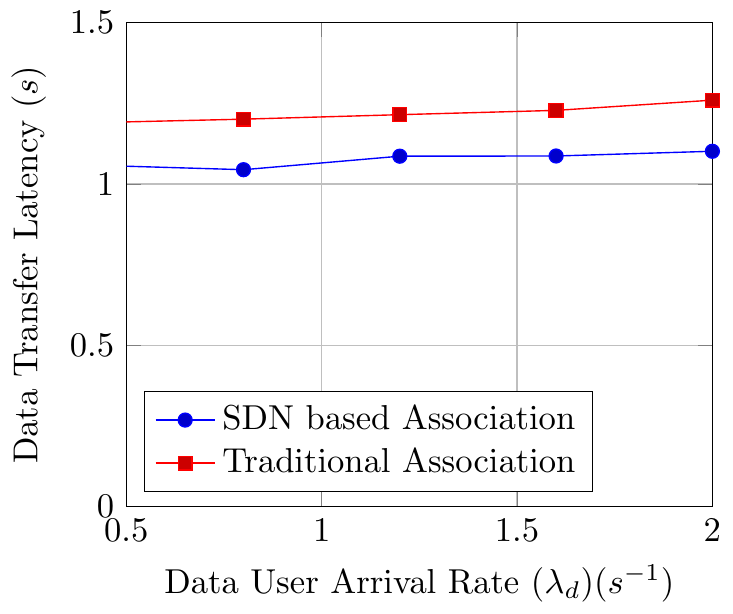}
		\caption{}
       \label{fig:system_latency}
	\end{subfigure} \\
    \begin{subfigure}{0.45\textwidth}
		\includegraphics{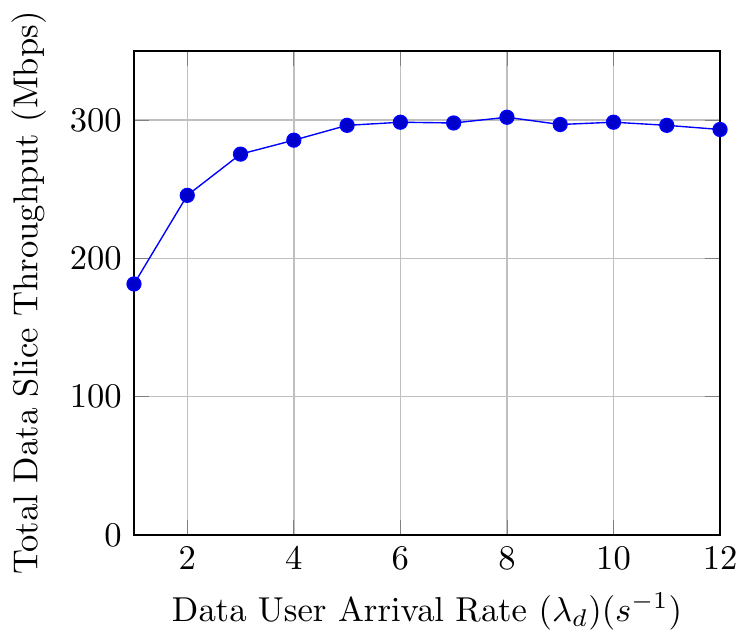}
		\caption{}
        \label{fig:slicing_throughput}
	\end{subfigure}
	\begin{subfigure}{0.45\textwidth}
		\includegraphics{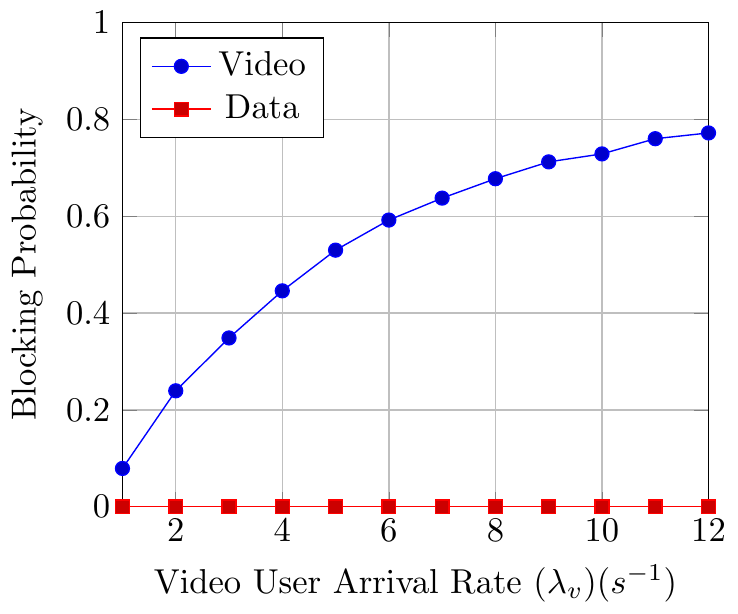}
		\caption{}
        \label{fig:blocking_probability}
	\end{subfigure}
\end{figure}

\par As illustrated in Figure~\ref{fig:system_throughput}, the system throughput for the SDN based multi-RAT is consistently better than that of the existing network. In the existing Multi-RAT network, a given RAT may not possess load information of other RATs. In the SDN based multi-RAT architecture, the presence of load information of all the constituent RATs at the controller improves association decisions, leading to improvement in total system throughput. We also observe that the end-to-end data packet latency as illustrated in Figure~\ref{fig:system_latency} is improved in the SDN framework. This is due to the fact that contention in WLAN increases with increased arrival rate, causing delays in packet transfers. However, since our architecture has a view of the WLAN load, data users could be associated with WLAN upto a threshold resulting in better performance. This scheme cannot be implemented in traditional networks due to the absence of a global view of the network.

\par In the second scenario, we illustrate that our framework supports network slicing and demonstrate that slice isolation can be ensured even when resources are common. We consider the same system model as in the previous case. The network is divided into two slices, one serving real-time video traffic (video-slice) and the other serving best-effort data traffic (data-slice). We envision that real time video traffic can be served by LTE RAT whereas best-effort data traffic can be served by both LTE and WLAN. Accordingly the video slice consists of LTE dBS alone and data slice consists of both LTE and WLAN dBSs. The data rate for real-time video users are assumed to be $400$kbps. We reserve $30\%$ of the LTE resources for real-time video, and the rest are reserved for data. The WLAN dBS is entirely reserved for data traffic. 

\par We measure the system throughput of the data-slice by varying the arrival rate of data users and maintaining constant video user arrival rate. As shown in Figure~\ref{fig:slicing_throughput}, the throughput of the data traffic increases upto the slice capacity and then remains constant. However, the throughput of the video slice remains unaffected. Similarly, if we increase the arrival rate of video users by keeping data user arrival rate constant, we can observe that the video traffic increases upto the slice capacity and then saturates. Video traffic which arrives when the slice capacity is reached is blocked without affecting the traffic in any other slice. This is illustrated in Figure~\ref{fig:blocking_probability} by the observed increase in blocking probability of video traffic with the growth in video user arrivals. 

\section*{Summary and Research Directions}
In this article, we have presented a brief overview of the architecture and prevalent issues in existing wireless multi-RAT networks. Details on some of the ongoing research and standardization activities towards the development of 5G wireless networks have also been provided. Further, we have proposed a novel SDN based network architecture for the unified control and management of multi-RAT networks. The architecture provides end-to-end network control, while ensuring scalability through creation of multiple logical networks over a single physical network. The advantages of this architecture have been demonstrated with the help of call flows and experimental results. 
\par Adopting a unified multi-RAT architecture opens up multiple avenues for further research. The presented architecture utilizes network slicing to meet the desired QoS requirements and achieve scalability. Since different slicing strategies are likely to have varying impact on network KPIs, strategies for slice creation and slice-to-service mapping are potential areas for future research. Policies for dynamic sharing of network entities across slices may also be explored. 
\par Design and standardization of interfaces between different network functions and development of efficient RRM algorithms are additional areas for research. The SDN based multi-RAT architecture abstracts RAT specific details and provides a common set of parameters to the applications. The parameter set required to satisfy QoS requirements for various services needs to be investigated. In conclusion, the article highlights the significant impact that the SDN may have on the design and development of wireless multi-RAT networks.

\section*{Acknowledgement}
This work has been supported by the Department of Telecommunications, Ministry of Communications, India as part of the indigenous 5G Test Bed project.

\bibliographystyle{ieeetr}
\bibliography{multirat}
\end{document}